\documentclass[preprint,12pt]{elsarticle}
\usepackage{amssymb}
\usepackage{graphicx}

\journal{Physics Letters B}

\begin{document}

\begin{frontmatter}



\title{Can Massive Gravitons be an Alternative to Dark Energy?}


\author{M. E. S. Alves, O. D. Miranda and J. C. N. de Araujo}

\address{INPE-Instituto Nacional de Pesquisas Espaciais - Divis\~ao
de Astrof\'isica,
\\Av.dos Astronautas 1758, S\~ao Jos\'e dos Campos, 12227-010 SP, Brazil\\}
\ead{alvesmes@das.inpe.br, oswaldo@das.inpe.br,
jcarlos@das.inpe.br}

\begin{abstract}
In this work, we explore some cosmological implications of the
model proposed by M. Visser in 1998. In his approach, Visser
intends to take in account mass for the graviton by means of an
additional bimetric tensor in the Einstein's field equations. Our
study has shown that a consistent cosmological model arises from
Visser's approach. The most interesting feature is that an
accelerated expansion phase naturally emerges from the
cosmological model, and we do not need to postulate any kind of
dark energy to explain the current observational data for distant
type Ia supernovae (SNIa).
\end{abstract}


\end{frontmatter}

\section{\label{sec:intr}Introduction}
The state of the art in cosmology has led to the following present
distribution of the energy densities of the Universe: 4\% for
baryonic matter, 23 \% for non-baryonic dark matter and 73 \% for
the so-called dark energy (see e.g. \cite{spergel06}). The dark
components (matter and energy) have been the focus of studies and
a lot of speculations in theoretical and observational
astrophysics.

In particular, the dark energy is very curious, not only for its
dominant relative energy density, but for its dynamical
consequences on cosmic expansion: it can make the Universe
accelerate. Phenomenologically, some equations of state have been
proposed in order to explain such a dark energy. The most common
example is the cosmological constant ($\Lambda {\rm CDM}$ model),
which implies on a constant vacuum energy density along the
history of the Universe. However, the quantum theory prediction
for the vacuum density is 120 orders of magnitude greater than the
observational value (see e.g. \cite{pad2003}).

Moreover, $\Lambda {\rm CDM}$ model is not satisfactory  because
it requires a large amount of fine tuning to produce cosmological
constant energy density dominant at recent epochs (see e.g.
\cite{Amendola07}).

Another possibility is a dynamical vacuum or quintessence. In
general, the quintessence models involve one \cite{Albrecht2000}
or two \cite{Bento2002} coupled scalar fields. In these models,
the cosmic coincidence, i.e., why dark energy started dominating
the cosmic evolution only so recently, has no satisfactory
solution and some fine tuning is required.

The Chaplygin gas is another example of dark energy fluid. Its
exotic equation of state can be derived from the Nambu-Goto action
for `D-Branes' moving on a (d+2) dimensional space-time
\cite{Gor2005}. The Chaplygin Gas has a dual behavior: in the past
it behaves like matter and in recent times like a cosmological
constant. So, it can represent dark matter and dark energy with
only one equation of state. However, at the time of structure
formation the influence of dark energy component is negligible and
matter clustering occurs in a similar way as in CDM model
\cite{Bert2005}.

More recently some authors (see, e.g., \cite{Amendola07}) have
studied the dark energy problem considering a quintessence model
with dark matter - dark energy interaction. However, as discussed
by Bertolami et al \cite{Bert07}, such a kind of coupling between
dark matter - dark energy produces a violation of the equivalence
principle (EP). Actually, a violation of the EP is reported to be
found in other dark energy models \cite{Füzfa07}.

Whatever the dark energy may be, it seems that physics beyond the
standard model is necessary.

On the other hand, several studies have been trying to bound the
mass of the graviton by many ways. From analysis of the planetary
motions in the solar system it was found that we must have $m_g <
7.8 \times 10^{-55}$g \cite{Tal1988} in order to respect the
accuracy of the observations with the newtonian potential. Another
bound comes from the studies of galaxy clusters, which gives $m_g
< 2 \times 10^{-62}$g \cite{gold1974}. Although this second limit
is more restrictive, it is considered less robust due to
uncertainties in the content of the Universe in large scales.

Studying rotation curves of galactic disks, de Araujo and Miranda
\cite{deAraujo2007} have found that $m_g \ll 10^{-59}g$ in order
to obtain a galactic disk with a scale length of $b\sim 10$ kpc.

Studying the mass of the graviton in the weak field regime Finn
and Sutton have shown that the emission of gravitational radiation
does not exclude a possible non null rest mass. They found the
limit $m_g < 1.4 \times 10^{-52}$g \cite{finn2002} analyzing the
data from the orbital decay of the binary pulsars PSR B1913+16
(Hulse-Taylor pulsar) and PSR B1534+12.

As can be seen, the graviton mass is not observationally excluded.
So, it is reasonable to ask if the consideration of mass for the
graviton plays some role in cosmology. Do cosmological models
exclude a non null graviton mass? Can massive gravitons affect the
cosmic dynamics? Is it possible to bound the graviton mass by
cosmological observation? Would such a bound be in accordance with
other observations? Considering mass for the graviton, do we need
to include dark energy in order to explain the cosmological data?

In order to look for answers to these questions, we have chosen to
include the graviton mass adopting the Visser's approach
\cite{vis1998}, where the graviton mass appears as an extra term
in the Einstein's equations. It is worth noting that the weak
field equations used by Finn and Sutton are the same that come
from the Visser's model when we use the linear approach.

An interesting result that comes out from this model is that the
gravitational waves present six polarization modes
\cite{paula2004} instead of the two usual polarizations obtained
from the General Relativity theory. So, if in the future we would
be able to identify the gravitational wave polarizations, we would
impose limits on the graviton mass by this way.

This paper is organized as follows: in Section 2 we briefly review
the Visser's approach. Section 3 is devoted to the description of
the cosmological model. In Section 4 we discuss how the age of the
Universe can constrain the value of the mass for the graviton.
Section 5 describes the evolution of the scale factor in the
`massive cosmology'. In Section 6 we show that a cosmological
scenario with massive gravitons can produce a phase of
accelerating expansion for the Universe. Section 7 presents a
comparison between $\Lambda$CDM and our cosmological model using
the luminosity distance. Finally, in Section 8 we present our
conclusions.

\section{\label{sec:two}The Field Equations}

The full action considered by Visser is given by \cite{vis1998}:
\begin{eqnarray}\label{fullaction}
I=\int d^4x\left[ \sqrt{-g}\frac{c^4R(g)}{16\pi G}+{\cal{L}}_{mass}(g,g_0) +{\cal{L}}_{matter}(g)\right].
\end{eqnarray}
where besides the Einstein-Hilbert lagrangian and the lagrangian
of the matter fields we have
\begin{eqnarray}
{\cal{L}}_{mass}(g,g_0) = \frac{1}{2}\frac{m_g^2c^2}{\hbar^2}\sqrt{-g_0}\bigg\{ ( g_0^{-1})^{\mu\nu} ( g-g_0)_{\mu\sigma}( g_0^{-1})^{\sigma\rho} \\ \nonumber
 \times ( g-g_0)_{\rho\nu}-\frac{1}{2}\left[( g_0^{-1})^{\mu\nu}( g-g_0)_{\mu\nu}\right]^2\bigg\},
\end{eqnarray}
where $m_g$ and $(g_0)_{\mu\nu}$ are respectively the graviton
mass and a general flat metric.

The field equations, which are obtained by variation of
(\ref{fullaction}), can be written as:
\begin{equation}\label{field-equations}
G^{\mu\nu} -\frac{1}{2}\frac{m_g^2c^2}{\hbar^2} M^{\mu\nu} = -\frac{8\pi G}{c^4}  T^{\mu\nu},
\end{equation}
where $G^{\mu\nu}$ is the Einstein tensor, $T^{\mu\nu}$ is the
energy-momentum tensor for perfect fluid, and the contribution of
the massive tensor to the field equations reads:
 \begin{eqnarray}\label{massive tensor}
   M^{\mu\nu} =  (g_0^{-1})^{\mu\sigma}\bigg[ (g-g_0)_{\sigma\rho}-\frac{1}{2}(g_0)_{\sigma\rho}(g_0^{-1})^{\alpha\beta}\\ \nonumber
   \times(g-g_0)_{\alpha\beta} \bigg](g_0^{-1})^{\rho\nu}    .
 \end{eqnarray}

Note that if one takes the limit $m_g\rightarrow 0$ the usual
Einstein field equations are recovered.

Regarding the energy-momentum conservation we will follow the same
approach of \cite{Narlikar1984} and \cite{Rastall1972} in such a
way that the conservation equation now reads \cite{Alves2006}:
 \begin{equation}\label{conservation}
   \nabla_\nu T^{\mu\nu} = \frac{m_g^2 c^6}{16\pi G \hbar^2} \nabla_\nu M^{\mu\nu},
 \end{equation}
since the Einstein tensor satisfies the Bianchi identities
$\nabla_\nu G^{\mu\nu} = 0$.

\section{\label{sec:three}Cosmology with massive gravitons and without dark energy}
For convention we use the Robertson-Walker metric as the dynamical
metric:
\begin{equation}\label{rwmetric}
ds^2=c^2dt^2-a^2(t)\left[ \frac{dr^2}{1-kr^2}+r^2(d\theta^2+\sin^2\theta d\phi^2)\right],
\end{equation}
where $a(t)$ is the scale factor. The flat metric is written in
spherical polar coordinates:
\begin{equation}\label{minkmetric}
ds_0^2=c^2dt^2-\left[ dr^2+r^2\left( d\theta^2+\sin^2\theta d\phi^2  \right)   \right].
\end{equation}

Using (\ref{rwmetric}) and (\ref{minkmetric}) in the field
equations (\ref{field-equations}) we get the following equations
describing the dynamics of the scale factor (taking $k=0$ for
simplicity):
 \begin{equation}\label{eqfried1}
   \left( \frac{\dot{a}}{a}\right)^2 + \frac{m_g^2 c^4}{4 \hbar^2}(a^2 - 1) = \frac{8\pi G}{3c^2} \rho
 \end{equation}
and
 \begin{equation}\label{eqfried2}
   \frac{\ddot{a}}{a}+\frac{1}{2}\left( \frac{\dot{a}}{a}\right)^2 + \frac{m_g^2 c^4}{ 8 \hbar^2}a^2(a^2-1) = -\frac{4\pi G}{c^2}  p  ,
 \end{equation}
where as usual $\rho$ is the energy density and $p$ is the
pressure.

From equation (\ref{conservation}) we get the evolution equation
for the cosmological fluid, namely:
 \begin{equation}
   \dot{\rho} + 3 H \left[  (\rho + p) + \frac{m_g^2 c^6}{32\pi G \hbar^2} (a^4 - 6a^2 + 3) \right] = 0,
 \end{equation}
where $H = \dot{a}/a$. Considering a matter dominated universe ($p
= 0$) the above equation gives the following evolution for the
energy density:
 \begin{equation}\label{rho-m-new}
   \rho = \rho_0\left( \frac{a_0}{a}\right)^3 - \frac{3m_g^2 c^6}{32\pi G \hbar^2} \left( \frac{a^4}{7} - \frac{6a^2}{5} + 1\right),
 \end{equation}
where $\rho_0$ and $a_0$ are the present values of the energy
density and the scale factor respectively. Note that in the case
$m_g \rightarrow 0 $ we obtain the usual Friedmann equations.

Now, inserting (\ref{rho-m-new}) in the modified Friedmann
equation (\ref{eqfried1}) we obtain the Hubble parameter:
 \begin{equation}\label{parHubMass}
   H^2(a)=H^2_0\left[\Omega^0_m \left( \frac{a_0}{a}\right)^3 + \frac{1}{140} \left(
   \frac{m_g}{m_H}\right)^2 \left( 7a^2 - 5a^4 \right)\right],
 \end{equation}
where the relative energy density of the $i$-component is
$\Omega_i=\rho_i/\rho_c$ ($\rho_c=3H^2c^2/8\pi G$ is the critical
density) where `$i$' applies for baryonic and dark matter. Here
$m_H=\hbar H_0/c^2$ is a constant with units of mass, which we
will call `Hubble mass' (as we will see later this constant is
very important in the present context).

\section{\label{sec:five}The Age of the Universe Constraining the Mass for the Graviton}

The equation (\ref{parHubMass}) for the present time ($a=a_0$)
gives:
\begin{equation}
\Omega_m^0 + \frac{1}{140} \left(\frac{m_g}{m_H}\right)^2 \left( 7a_0^2 - 5a_0^4 \right) =1.
\end{equation}

Solving this equation to $a_0=a_0(m_g)$ we get:
\begin{equation}\label{ramg}
a_0 = \sqrt{\frac{7}{10}} \left\lbrace 1+ \left[ 1-\frac{200}{7} \left( \frac{m_H}{m_g}\right)^2\left( 1-\Omega^0_m\right) \right] ^{\frac{1}{2}}\right\rbrace ^{\frac{1}{2}}.
\end{equation}
Thus, if we use (\ref{ramg}) in the equation (\ref{parHubMass}) we
have the dynamical equations described in terms of the free
parameters $\Omega_m^0$, $H_0$ and $m_g$.

Note that in order to have real values for $a_0$, the term in
brackets in (\ref{ramg}) must satisfy the relation:
\begin{equation}
1-\frac{200}{7} \left( \frac{m_H}{m_g}\right)^2\left( 1-\Omega_m^0\right) > 0,
\end{equation}
which lead us to a lower limit for the graviton mass in our model:
\begin{equation}\label{inferior bound}
m_g > \sqrt{\frac{200}{7}\left( 1-\Omega_m^0\right)}~m_H.
\end{equation}

If we take, for example, $\Omega_m^0=0.27$ we have:
\begin{equation}
m_g > 4.57 m_H.
\end{equation}

So, the Hubble mass, $m_H$, establishes the magnitude of the
expected value of the graviton mass. If we convert this lower
limit in a upper limit to the Compton wavelength of the graviton
we have:
\begin{equation}
\lambda_g < 0.22 \frac{c}{H_o}.
\end{equation}

\begin{figure}[!ht]
\centering
\includegraphics[width=90mm]{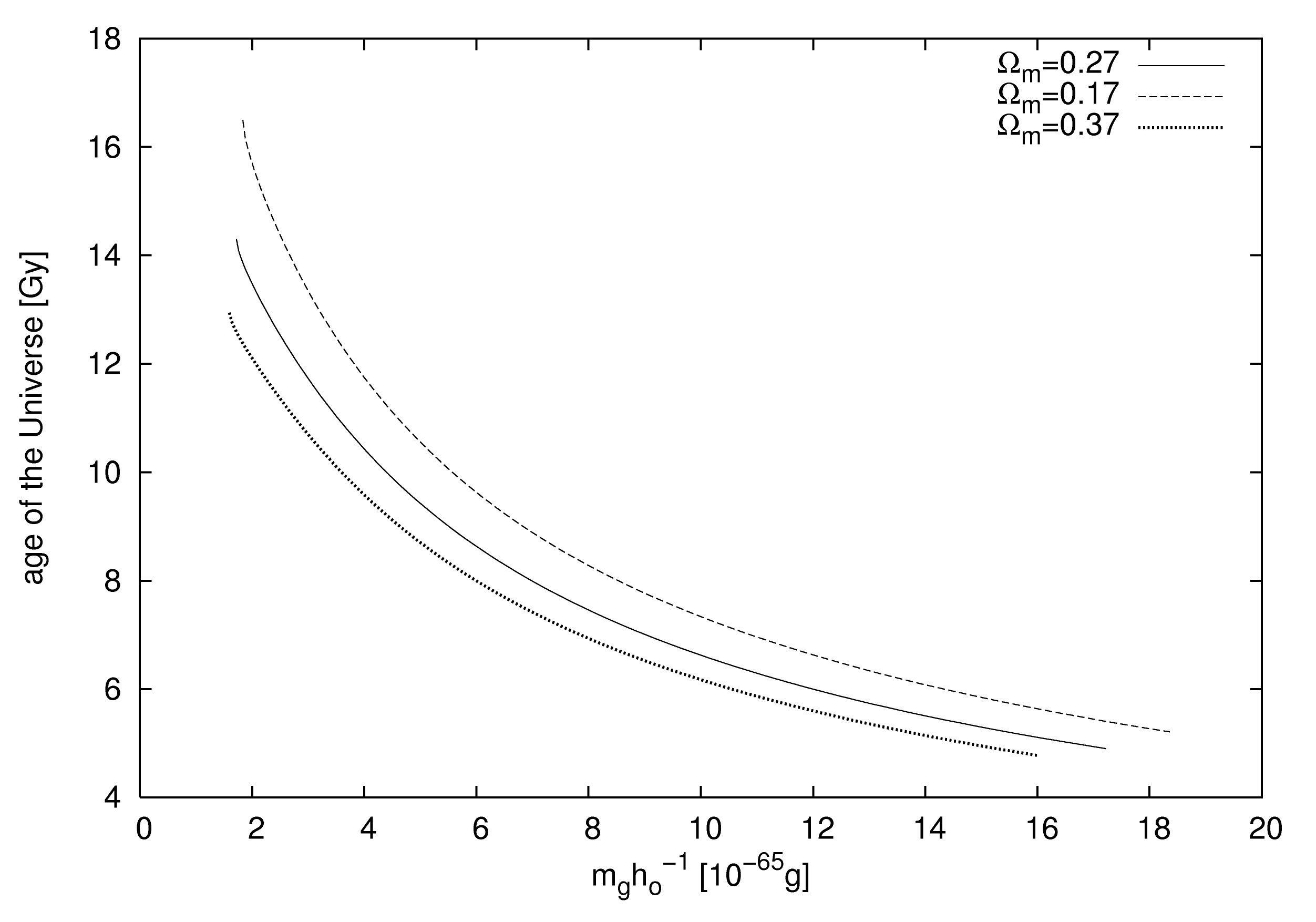}
\caption{Relation between the age of the Universe and the mass for the graviton for different values of $\Omega_m$.}
\label{fig1}
\end{figure}

Thus, if the graviton rest mass is non null, its corresponding
Compton wavelength must be lower than the observable horizon. But
once the Compton wavelength is associated with the range of the
interaction, this tell us that the contribution of the mass term
can be relevant in the Universe at the present time.

A note about the lower bound on the graviton mass. Since we are
considering only matter in the context of a massive gravity
theory, the lower bound on the graviton mass is a direct
consequence of the fact that we have $\Omega_{total}\simeq 1$
today. Once matter contributes only with $27\%$ of the total
relative energy density, and the radiation contribution is
negligible, we could not have a zero graviton mass without
considering another kind of dominating component, but such idea is
not the aim of the present paper.

Once we have an analytical description of the dynamics in this
scenario we can impose limits on the parameters in many ways. One
of them can be the age of the Universe.

The time scale to the age of the Universe is related to the Hubble
parameter, namely, $t_H=H_0^{-1}$, which is about $13$ Gy as given
by the current observational data \cite{spergel06}.

So, we hope that a consistent cosmological model can give ages of
this order, which contemplates the age of the oldest stars and the
time to structure formation.

In order to calculate the age of the Universe in our model we just
solve the integral:
\begin{equation}\label{lookbacktime}
t_U=\int_0^{a_0} \frac{da}{aH(a)}.
\end{equation}

Substituting (\ref{parHubMass}) in (\ref{lookbacktime}) we obtain
the age of the Universe as shown in Figure \ref{fig1} to some
values of $\Omega_m^0$. We can see that the age of the Universe is
very closely related to the mass of the graviton. This gives us a
very restrictive upper limit. The effect of taking different
values of $\Omega_m^0$ is to shift the curve upward or downward,
but the qualitative behavior is the same in all cases.

If we take, for example, $t_H=13$ Gy and $\Omega_m^0=0.27$ we
obtain the limit:
\begin{equation}
m_gh_0^{-1} \lesssim 2.24\times 10^{-65}~{\rm{g}}~,
\end{equation}
where the Hubble parameter is given by
$H_0=100h_0~kms^{-1}Mpc^{-1}$ and $h_0$ is a dimensionless
constant, which parameterizes the uncertainty in the measurement
of $H_0$. This limit for $m_g$ is about $10$ orders of magnitude
more restrictive than the limits obtained from the orbital motion
in solar system and about $3$ orders more restrictive than the
inferences from galaxy clusters.

Considering the lower limit obtained previously, the mass of the
graviton in this model must be in the interval:
\begin{equation}\label{limite2}
1.73 \times 10^{-65}~{\rm{g}}~<~m_gh_0^{-1}~\lesssim~2.24\times 10^{-65}~{\rm{g}}~,
\end{equation}
for $\Omega_m^0=0.27$. This upper limit depends on the value taken
for the age of the Universe, while both upper and lower limits
depend on the value of $\Omega_m^0$.

\section{Past and Future}

\begin{figure}[!ht]
\centering
\includegraphics[width=90mm]{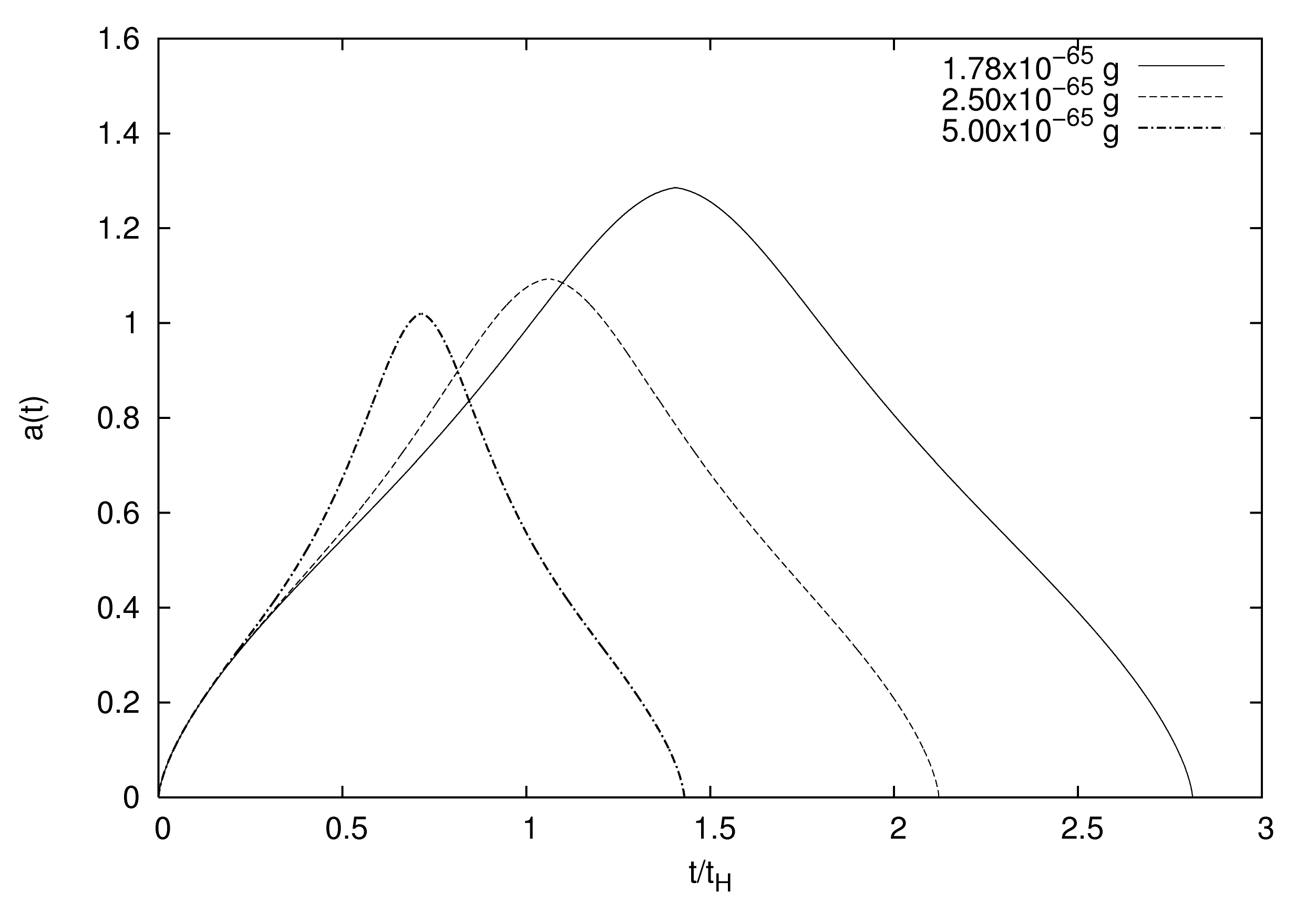}
\caption{Evolution of the normalized scale factor
with time (in units of the Hubble time) for different values of
$m_g$.} \label{fig2}
\end{figure}

One can determine how the Universe evolves by integrating equation
(\ref{parHubMass}), which yields $a(t)$ as shown in Figure
\ref{fig2}.

In the past and in the present the curve $a(t)$ behaves like the
evolution of the Friedmann models with a cosmological constant.
However, the future is drastically different. The massive term
contributes in such a way that in the past it generates a
repulsion, becoming attractive in the future and leading the
Universe to the so called big crunch. Moreover, although the
evolution is qualitatively the same for any value of the graviton
mass, it is notable the strong dependence between the time of
evolution and $m_g$. This emphasizes again the restricted values
to $m_g$ which we are considering.

\begin{figure}[!ht]
\centering
\includegraphics[width=90mm]{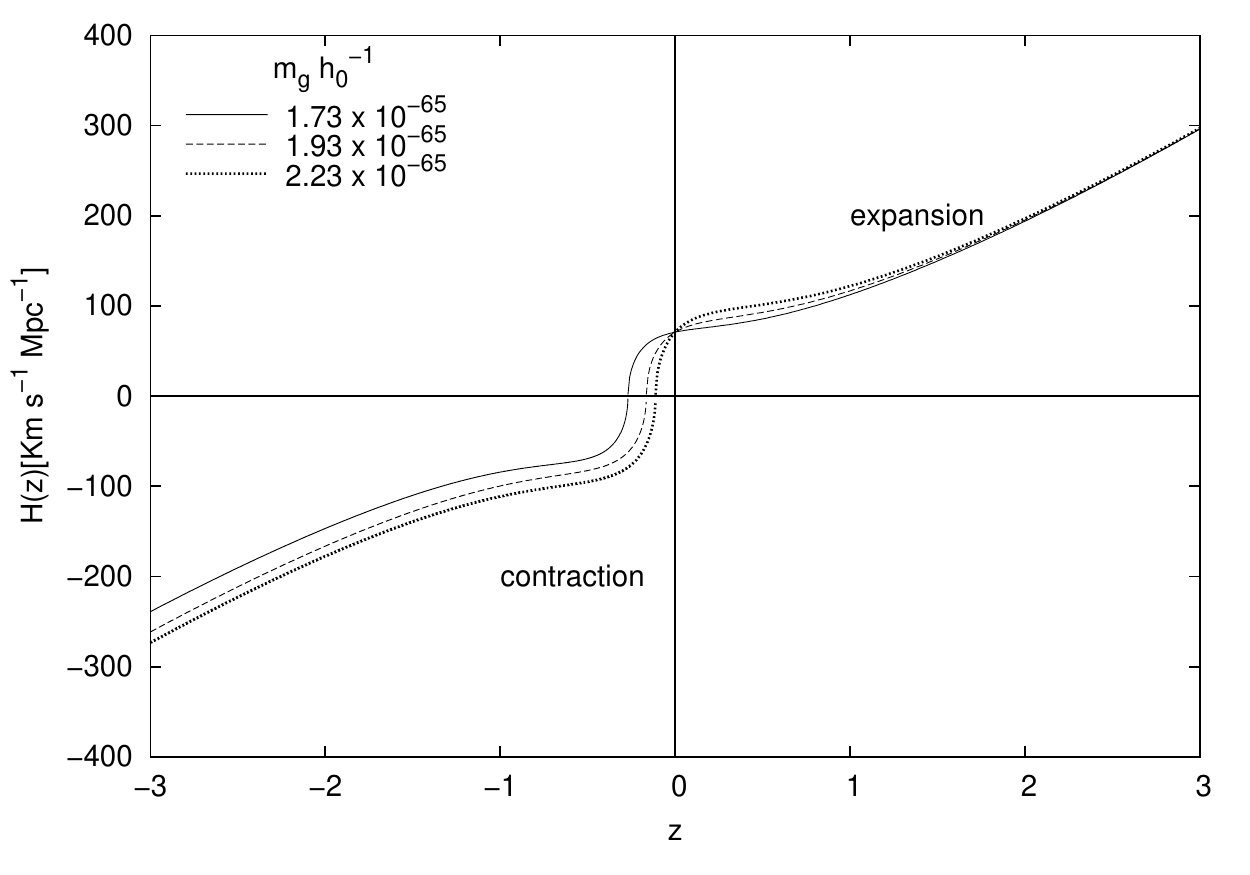}
\caption{Evolution of the Hubble parameter as a
function of the redshift for different values of $m_g$. We
take $\Omega_m^0=0.27$.} \label{fig3}
\end{figure}

In the Figure \ref{fig3} we can see the evolution of the Hubble
parameter. In the past the function behaves like the Friedmann
models and in the future $H(z)\rightarrow0$. This shows us that
the massive contribution becomes important in the late time for
the history of the Universe. In fact, if we take $a \rightarrow0$
in equation (\ref{parHubMass}) we see that the massive term is
negligible when compared to the radiation and matter contribution.
That is very important in order to have no change in the expansion
rate of the Universe, for example, in the nucleosynthesis era (a
detailed study of this issue will appear elsewhere).

When the scale factor reaches its maximum value (the turning point
$z_{turn}$), the Hubble parameter goes to zero $H(z_{turn})=0$.
The smaller $m_g$ is, the greater $z_{turn}$ is. This indicates
the relation between the turning point and the range of the
gravity interaction, once the Compton wavelength is inversely
proportional to the mass of the particle.

\section{Accelerating Universe?}

In order to study the evolution of the second derivative of the
scale factor we will use the definition of the dimensionless
desaccelerating parameter:
\begin{equation}
q\equiv -\frac{\ddot{a}a}{\dot{a}^2}~.
\end{equation}

Applying this to the equations (\ref{eqfried1}) and
(\ref{eqfried2}) we get:
 \begin{equation}\label{desacelerating}
   q(a)=\frac{H^2_0}{H^2}\left[ \frac{1}{2}\Omega_m^0\left(\frac{a_0}{a}\right)^3+\left( \frac{m_g}{m_H}\right)^2 \left(\frac{3a^4}{14}-\frac{a^2}{5}\right) \right].
 \end{equation}

\begin{figure}[!ht]
\centering
\includegraphics[width=90mm]{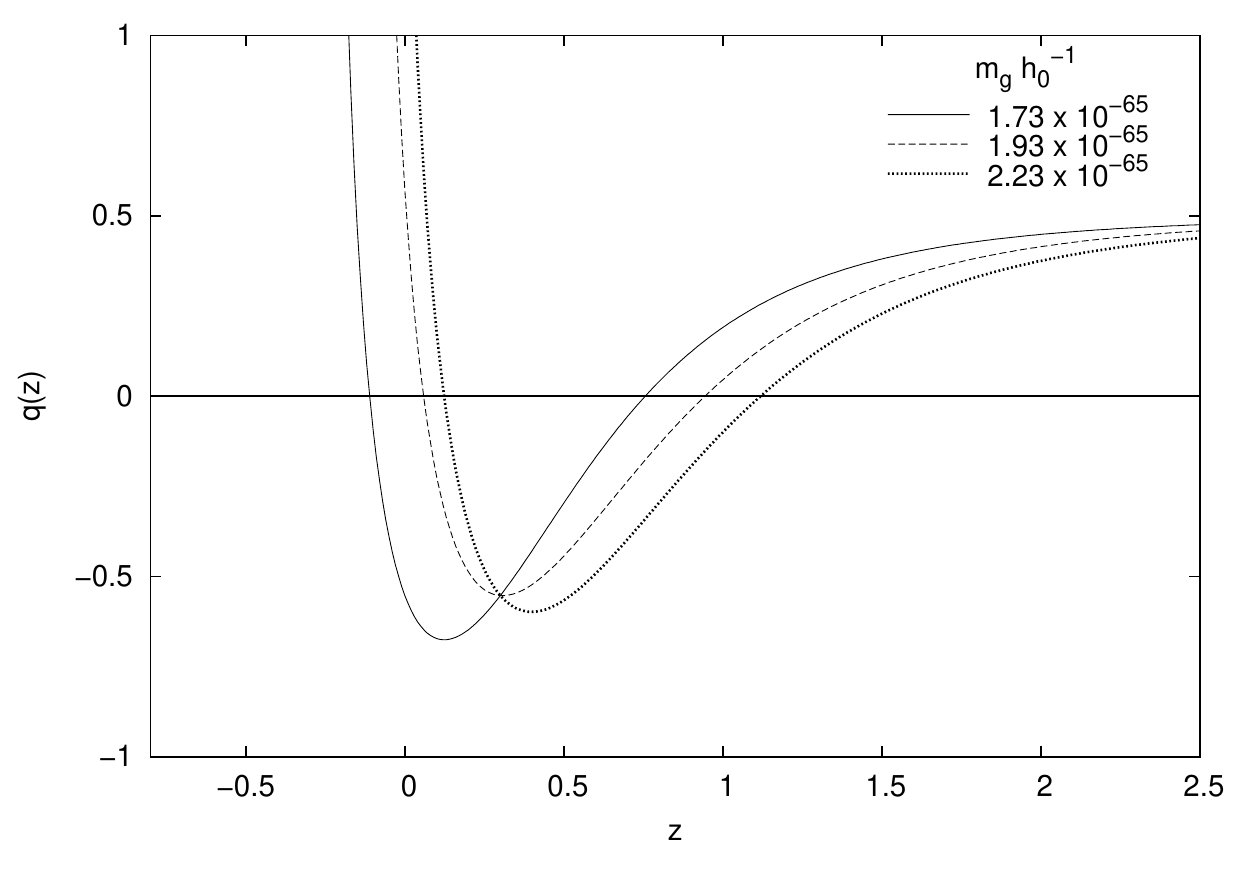}
\caption{Evolution of the desaccelerating parameter $q(z)$ for different values of $m_g$. We take $\Omega_m^0=0.27$}
\label{fig4}
\end{figure}

The evolution of the desaccelerating parameter as a function of
the redshift is shown in Figure \ref{fig4}.

Note that an accelerating expansion at the present time is closely
related to the value of the graviton mass. In the past, as
expected, the Universe has a desaccelerated phase. The fast growth
of $q(a)$ in the future is explained by the reversion in the
expansion as discussed above. When the Universe is contracting,
the values of $q(a)$ are the same as in the expansion phase, which
lead us to conclude that the Universe has two accelerating phases
in this model: the first when it expands and the second when it
contracts.

In order to verify the dependence between the present value of the
desaccelerating parameter ($q_0$) and the value of $m_g$ we just
set $a=1$ in the equation (\ref{desacelerating}), giving:
\begin{equation}
q_0=\frac{1}{2}\Omega_m^0+\left( \frac{m_g}{m_H}\right) ^2\left( \frac{3a_0^4}{14}-\frac{a_0^2}{5}\right).
\end{equation}

\begin{figure}[!ht]
\centering
\includegraphics[width=90mm]{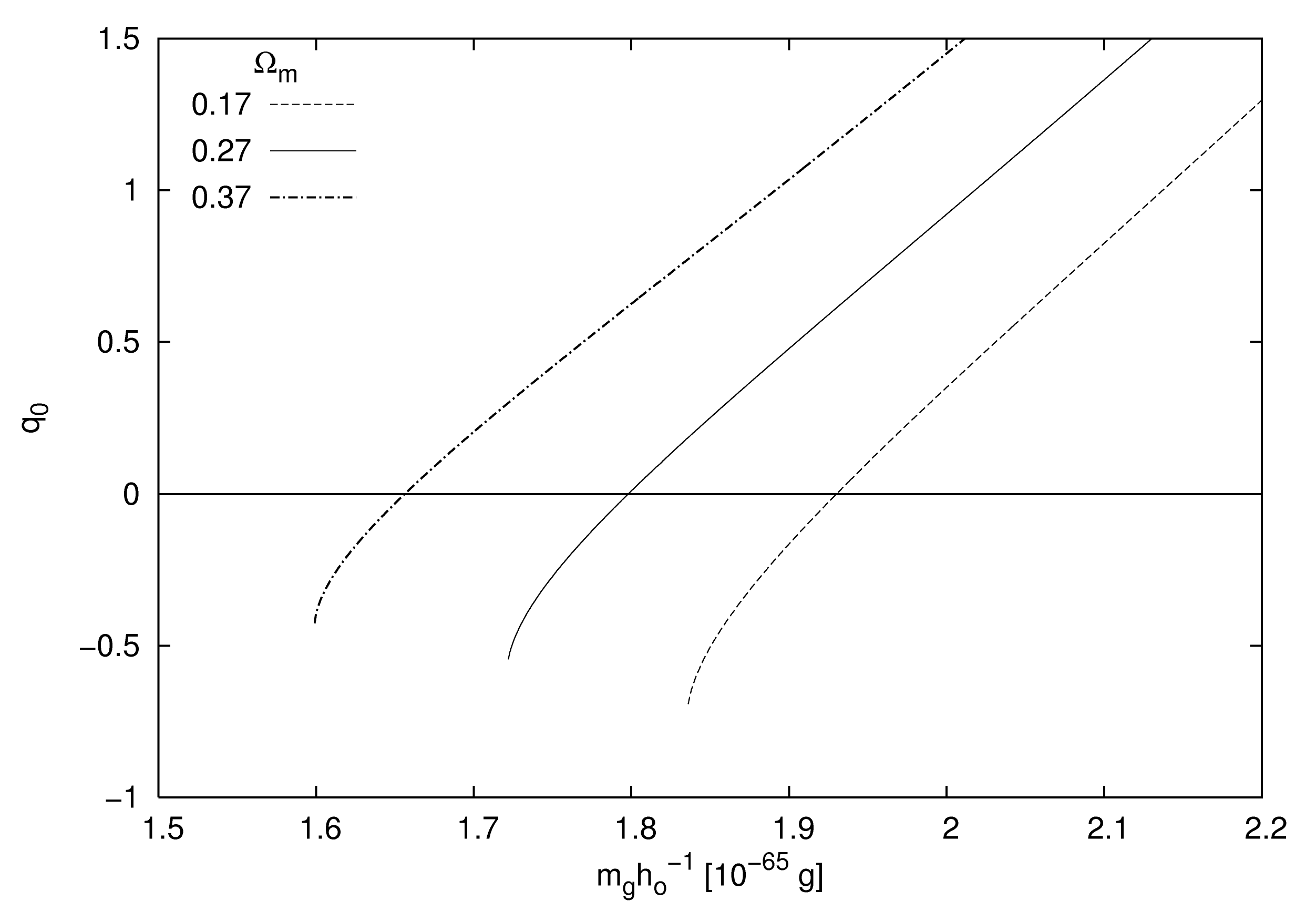}
\caption{Relation between the current value of the desaccelerating parameter and the graviton mass for different values of $\Omega_m$.}
\label{fig5}
\end{figure}

The curve $q_0 \times m_g$ is shown in Figure \ref{fig5}. If we
assume $\Omega_m^0=0.27$, for example, we have a present
accelerating expansion for $m_gh_0^{-1} < 1.80 \times 10^{-65}$g.

It is interesting to observe that this result is compatible with
older Universes and greater lifetimes, which indicates that the
most probable value for $m_g$ would be closer to the lower limit
(see equation (\ref{limite2})).

\section{The Luminosity Distance}

In order to test this cosmological model, we calculate the
luminosity distance and compare it with the $\Lambda$CDM best fit
of the type Ia supernovae.

The luminosity distance to a flat Universe ($k=0$) is given by
(see, e.g. \cite{Sahni2000}):
\begin{equation}\label{luminosity dist}
d_L(z)=(1+z)cH_0^{-1}\int^z_0\frac{dz^{\prime}}{h(z^{\prime})},
\end{equation}
where $h(z)=H(z)/H_0$ and $(1+z)=a^{-1}$.

In the $\Lambda$CDM model the equation (\ref{luminosity dist})
gives us:
\begin{equation}\label{lum-dist-lcdm}
d_L^{\Lambda CDM}=\frac{(1+z)c}{H_0}\int^z_0\frac{dz^{\prime}}{\sqrt{\Omega_m^0(1+z^{\prime})^3+\Omega_\Lambda^0}}~.
\end{equation}

To the massive model we can combine (\ref{parHubMass}) with
(\ref{luminosity dist}) to obtain:
\begin{equation}\label{lum-dist-mass}
d_L^{mass}=(1+z)cH_0^{-1}\int^z_0dz^{\prime}\sqrt{\Pi(z^{\prime})}~,
\end{equation}
where:
\begin{equation}\label{pi-cdm}
\Pi(z^{\prime})\equiv \frac{(1+z^{\prime})^4}{\Omega^0_m(1+z^{\prime})^7-M(z^{\prime})}~,
\end{equation}
and
$$
M(z^{\prime}) \equiv (m_g/m_H)^2\left[ a_0^4/14-(a_0^2/10)(1+z^{\prime})^2\right]~.
$$
If we use the relation between $a_0$ and $m_g$, given by
(\ref{ramg}), the luminosity distance in our model has the same
number of free parameters as in the $\Lambda$CDM model.

\begin{figure}[!ht]
\centering
\includegraphics[width=90mm]{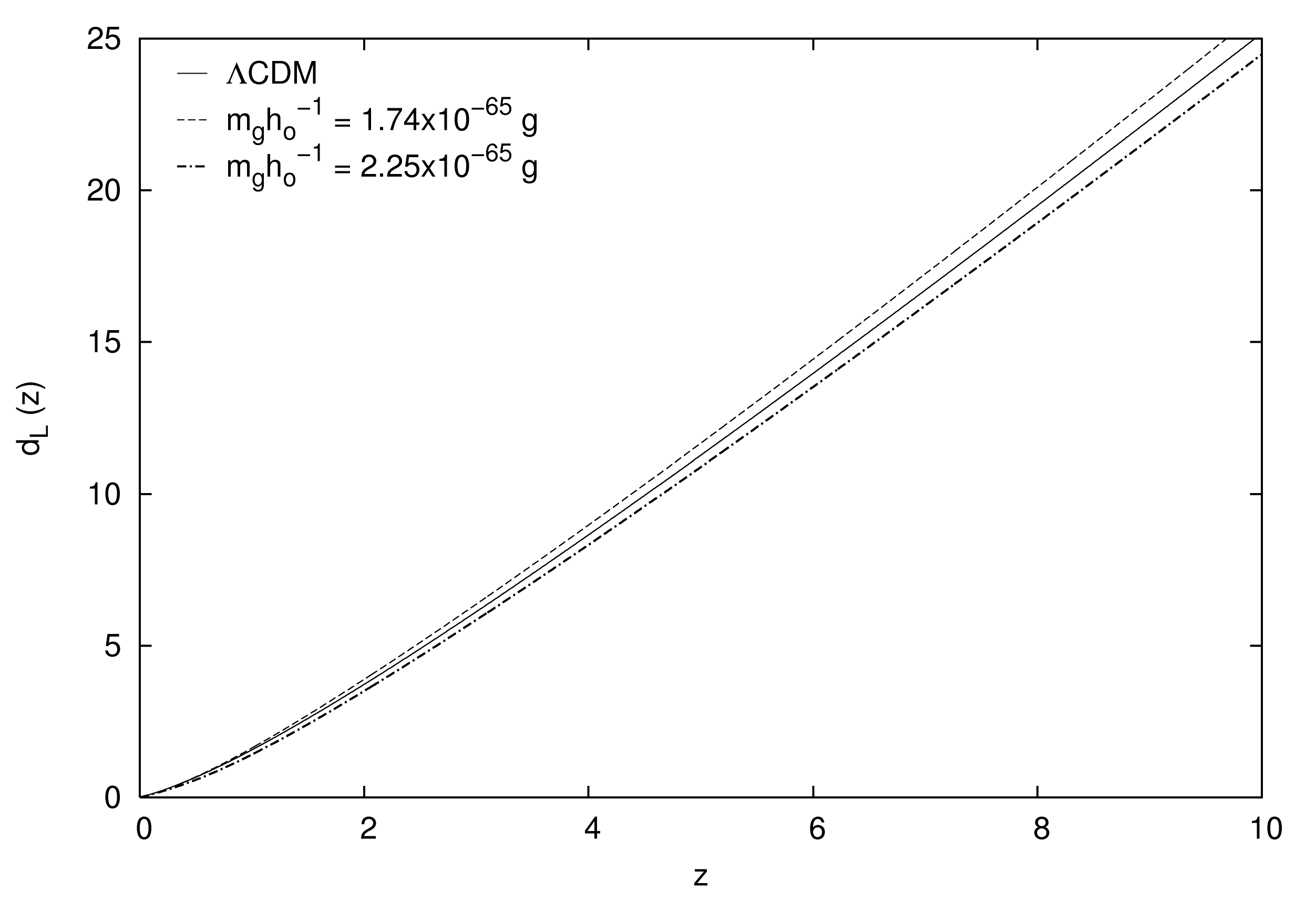}
\caption{Luminosity distance (in units of $cH_0^{-1}$) as a function of the redshift for a $\Lambda$CDM model and for the massive model.}
\label{fig6}
\end{figure}

\begin{figure}[!ht]
\centering
\includegraphics[width=90mm]{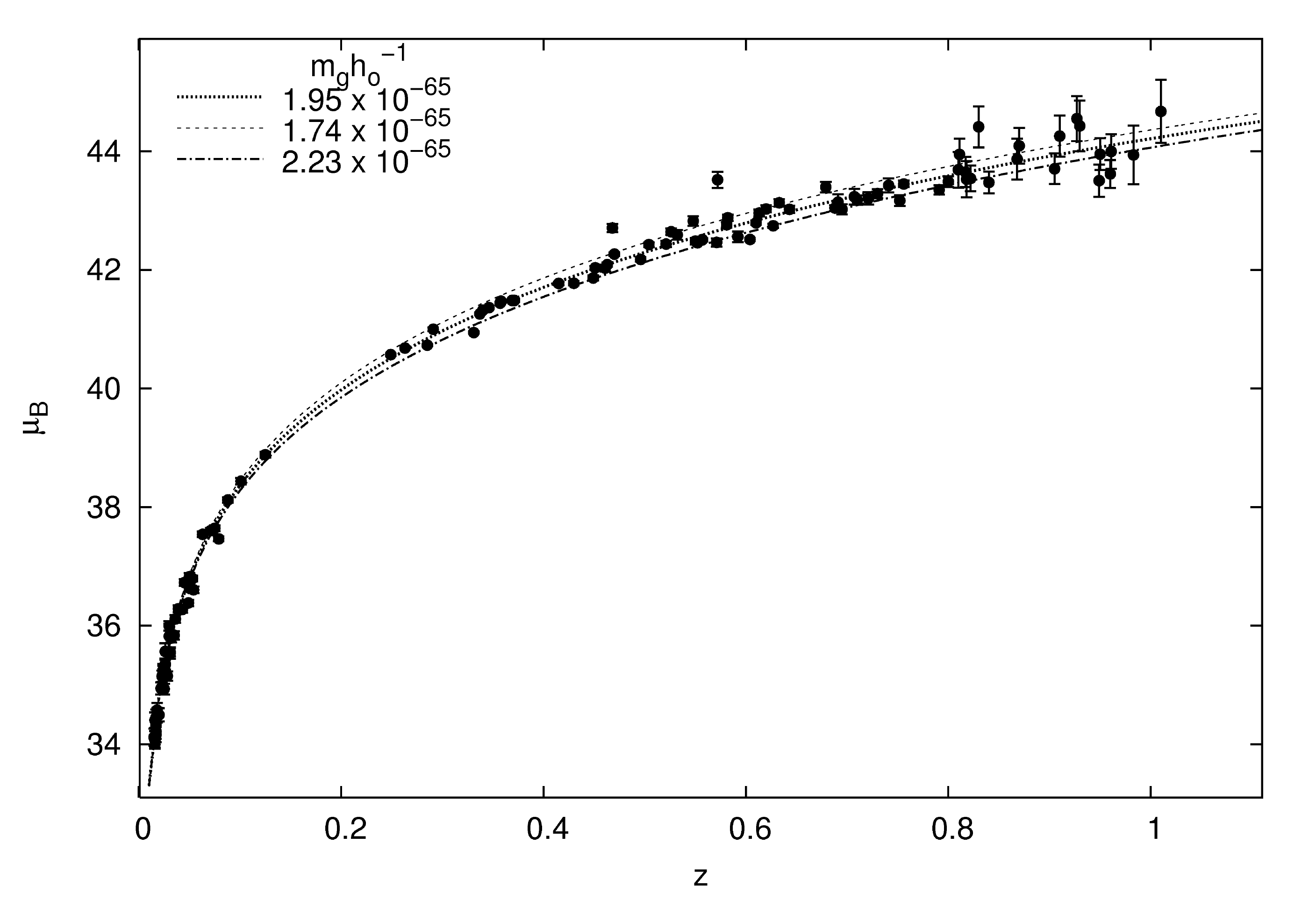}
\caption{Distance modulus as a function of the redshift to the massive model compared with the SNIa data from Astier et al. (2006).}
\label{fig7}
\end{figure}

In Figure \ref{fig6} we plot some values of the graviton mass for
the limits given by (\ref{limite2}). We also plot the curve for
the $\Lambda$CDM model, which is the best fit for the most recent
observational data from the recession of type Ia Supernovae, i.e.,
$\Omega_m^0=0.27$ and $\Omega_{\Lambda}^0=0.73$ \cite{astier06}.
The comparison between the model and the SNIa data can be seen in
Figure \ref{fig7}, where we use the definition of the distance
modulus:
\begin{equation}
\mu \equiv m-M=5\log(d_L)-5+A~,
\end{equation}
where $m$ and $M$ are the apparent and absolute magnitude
respectively and $A$ is the absorption in magnitude.

As can be seen, there are not significant differences in the
curves $d_L\times z$, when we change the graviton mass in that
range we have considered. Furthermore, we can give a value to
$m_g$ which describes exactly the same curve given by the
$\Lambda$CDM model, this value is around $m_gh_0^{-1}=1.95 \times
10^{-65}$ g. From Figure (\ref{fig5}) we see that for this value
of the graviton mass we have a positive desaccelerating parameter
in the present ($z = 0$), so we can fit the observational data for
SNIa without a present acceleration in the expansion of the
Universe, although we have an inevitable accelerating expansion
era in the near past.

Again, we would like to emphasize that there are no significant
changes in the curve $d_L\times z$ when we use other values of
$m_g$, which are in the established limit. In such a way there are
some possible values of $m_g$, which give current acceleration and
other that give desacceleration. For example, taking the values
used in the Figure \ref{fig7}, the lower one ($1.74\times
10^{-65}$ g) gives current desacceleration ($q_0>0$), and the
greater one ($2.23 \times 10^{-65}$ g) gives a current cosmic
acceleration ($q_0<0$).

In view of the above results we may conclude that it is not
possible to decide between the massive model and $\Lambda$CDM
model only by the type Ia Supernovae observations, in particular
for the small redshifts in which the Supernovae have been observed
($z \lesssim 1$).

\section{Conclusions}
We have shown that within the context of a classical gravity
theory with massive gravitons, we can obtain a consistent
cosmological model which has acceptable values for the age of the
Universe, furthermore, it can fit the present cosmological SNIa
data without any kind of dark energy.

The possible values for the graviton mass in this theory are in
accordance with all established limits.

We identify that there is degeneracy between the $\Lambda$CDM
model and the massive model when we analyze the luminosity
distance versus redshift. Such a degeneracy could be removed, in
principle, by other tests of the massive model such as structure
formation or analyzing the power spectrum of the cosmic microwave
background radiation (CMBR).

In future works we will study in detail the nucleosynthesis era
within the context of the massive model, as well as to provide a
statistical approach to the analysis of the SNIa data within this
model.

Once the graviton mass is introduced via an effective tensor as
given by (\ref{field-equations}), we hope that the primordial
graviton production would be different from those models which
consider general relativity. So, the future detection of
primordial gravitational waves will provide a way to test this
alternative gravity theory.

It is worth stressing that Gabadadze and Gruzinov (2005) have
analyzed the instabilities of the background and ghosts produced
by massive gravitons in 4D Minkowski space-time. They conclude
that a natural way to account for a massive graviton on flat space
is to invoke theories with extra dimensions. However,  Visser's
model is truly continuous with general relativity (GR) in the
limit of vanishing graviton mass. Together with Pauli-Fierz (PF)
massive term, Visser's theory is the simplest attempt to
incorporate mass for the graviton in a ghost-free manner.
Moreover, the van Dam-Veltman-Zakharov discontinuity
(vDVZ)\cite{van Dam1970} present in the PF term can be
circumvented in Visser's model by introducing a non-dynamical
flat-background metric \cite{will2006}.

Thus, Visser's model can help us to understand the influence of
the mass for the graviton in cosmology and maybe it could shed
some light on the following question: what is the dark energy?

\section*{Acknowledgments}
The authors thank Prof. Jos\'e A. de Freitas Pacheco for
clarifying some important points and for a critical reading of the
manuscript. We also thank Prof. Jos\'e A.S. Lima for helpful
discussions. MESA would like to thank the Brazilian Agency FAPESP
for support (grant 06/03158-0). ODM and JCNA would like to thank
the Brazilian agency CNPq for partial support (grants
305456/2006-7 and 303868/2004-0 respectively).

\end{document}